# A Comparison of Anodic TiO$_2$ Nanotube Membranes used for Front-side Illuminated Dye-Sensitized Solar Cells

by Fatemeh Mohammadpour,[1] Mahmood Moradi,[1] Gihoon Cha,[2] Seulgi So,[2] Kiyoung Lee,[2] Marco Altomare[2] and Patrik Schmuki[2,3,*]

[1] Department of Physics, Faculty of Science, University of Shiraz, Shiraz 71454, Iran

[2] Department of Material Science and Engineering, WW4-LKO, University of Erlangen-Nuremberg, Martensstrasse 7, D-91058 Erlangen, Germany

[3] Department of Chemistry, King Abdulaziz University, Jeddah, Saudi Arabia

* Corresponding author: Tel.: +49-9131-852-7575, Fax: +49-9131-852-7582
Email: schmuki@ww.uni-erlangen.de

# A B S T R A C T

In the present work we compare TiO$_2$ nanotube lift-off strategies for the construction of front-side illuminated dye-sensitized solar cells (DSSCs). Anodic nanotube layers were detached from the metallic back contact by using different techniques and transferred onto an FTO substrate. We show that if we use an optimized potential step treatment to fabricate membranes, DSSC cell efficiencies can be significantly increased ($\eta > 8\%$). This improved efficiency is ascribed to higher specific dye-loading and enhanced electron transport properties of optimally fabricated TiO$_2$ nanotube membranes.

## 1. Introduction

In the last decades anodic $TiO_2$ nanotubes (NTs) have been widely used as photo-anodes in dye-sensitized solar cells (DSSCs) [1-4], this owing to their highly ordered geometry [5-7], superior optical and electronic properties [8-10], and expected one-dimensional electron transport [11-14].

When dye sensitized $TiO_2$ scaffolds are used in DSSCs, typically a front-side illumination configuration of the photo-anode is of advantage for reaching a high cell efficiency; this because the photo-anode is irradiated through the FTO slide and light losses, namely by light absorption in the electrolyte and in the Pt counter electrode, can be minimized.

In the case of nanotube layers, a front-side configuration can be achieved by a detachment of the aligned nanotube layer from Ti substrate as free-standing membrane that then is transferred to a conductive glass substrate (FTO). Different approaches have been reported to accomplish successful detachment of the tube arrays from the metal substrate, and the transferred $TiO_2$ nanotube layers have been explored in DSSCs in various configurations (tube bottom up/down and open/close tube bottom) [15-21]. However, a direct comparison of various results is difficult, as solar cell efficiency measurements are generally made with different type of cell constructions. The present work thus investigates different approaches and compares the feasibility to improve DSSC efficiencies for transferred membrane layers, in order to optimize the membrane fabrication approach.

## 2. Experimental

Titanium foils (0.125 mm thick, 99.7 % purity, Advent) were degreased by sonication in acetone, ethanol and deionized water, and finally dried in a nitrogen stream. All anodization experiments to grow self-assembled $TiO_2$ nanotube layers were performed in ethylene glycol-based electrolyte with 3 vol% water and 0.15 M $NH_4F$.

For producing free-standing membranes, two different strategies were used to detach the tube arrays from the metallic substrate that are namely "re-anodization" and a "potential shock" technique.

For the re-anodization method, $TiO_2$ nanotubes (12 µm long, grown by 25 min-long anodizations at 60 V) were crystallized by annealing in air at 350 °C for 1h. Then, these crystalline tubes were re-anodized in the same experimental conditions. This resulted in the formation of a second layer of amorphous NTs beneath the crystalline tubes. These amorphous tubes can preferentially be dissolved in aqueous solutions of HF (0.07 M) at 30 °C or of $H_2O_2$ (30%) at room temperature so that the crystalline $TiO_2$ nanotube arrays are detached from the Ti substrates.

$TiO_2$ nanotube membranes were also fabricated by a "potential shock" strategy. For this, $TiO_2$ nanotubes were grown at 60 V for 1 h, that is to a layer thickness of *ca*. 20 µm. Then, a potential shock was applied that led first to acceleration of the tube growth and then to breaking of the tubes by inducing stress and breakdown phenomena [22,23].

After detachment, the tube membranes were transferred to FTO substrates (7 Ω $cm^{-1}$) that were previously coated with a 2 µm-thick $TiO_2$ nanoparticle film (Ti-Nanoxide HT, Solaronix) deposited by a Doctor Blade method. After drying in air for about 20 min, the photo-anodes were annealed at 500 °C in air for 1 h, with heating/cooling rate of 30 °C $min^{-1}$, by using a Rapid Thermal Annealer (Jipelec JetFirst100).

The as-prepared photo-anodes were then immersed into a 300 µM dye solution (D719, Everlight, Taiwan) at 40 °C for 24 h. After dye-sensitization the samples were rinsed with acetonitrile to remove the non-chemisorbed dye and dried in $N_2$ stream. Then the photo-anodes were sandwiched with a Pt coated FTO glass as a counter electrode by using a hot-melt spacer (25 µm, Surlyn, Dupont). The electrolyte (Io-li-tec, ES-0004) was introduced in the interspace between the two electrodes of the DSSCs.

For morphological characterization, a field-emission scanning electron microscope (FE-SEM, Hitachi S4800) was used. X-ray diffraction analysis (XRD) was performed with an X'pert Philips MPD diffractometer with a Panalytical X´celerator detector using graphite monochromized Cu Kα radiation (λ = 1.54056 Å).

The current voltage characteristics of the DSSCs were measured under simulated AM 1.5 illumination provided by a solar simulator (300 W Xe lamp with optical filter, Solarlight) applying an external bias to the cell (from - 50 to + 900 mV) and measuring the generated photocurrent with a Keithley model 2420 digital source meter. Step size and holding time were 23.75 mV and 100 ms, respectively. The active (irradiated) area of the solar cells was defined by the 0.2 $cm^2$-sized opening of the Surlyn seal and scattering background beneath the solar cells was used.

Intensity modulated photovoltage spectroscopy (IMPS) measurements were carried out using modulated light (10 % modulation depth) from a high power green LED (λ = 530 nm). The modulation frequency was controlled by a frequency response analyzer (FRA, Zahner). The light intensity incident on the cell was measured using a calibrated Si photodiode.

Dye loading on the $TiO_2$ nanotube membranes was measured by immersing the dye-sensitized nanotube layers in 5 mL of a 10 mM NaOH aqueous solution for 30 minutes. Then the absorption of the solutions was measured by UV-Vis Spectrophotometer (Lambda XLS+, Perkin Elmer).

## 3. Results and discussion

TiO$_2$ NTs used in the present work were formed by anodization of Ti foil at 60 V for 25 min in ethylene glycol-based electrolyte (containing 3 vol% water and 0.15 M NH$_4$F). This results in 12 μm-thick TiO$_2$ nanotube layers with individual tube diameter of 80-100 nm, as shown in Fig. 1A.

As outlined in the experimental section, to detach the nanotube layers two different approaches were explored that are the re-anodization and the potential shock. The former approach follows a treatment common to Al$_2$O$_3$ membranes or reported to detach TiO$_2$ nanotube layers from Ti [15-17,24]. Layers fabricated by re-anodization could indeed be successfully lifted off as entire membranes by anodizing for a second time NT layers that were previously crystallized (first and second anodization steps were carried out in same experimental conditions). In particular, we found that the use of HF led to detachment of entire tube layers that showed complete close tube bottoms (Fig. 1B). Differently, membranes with partially open tube bottom could be fabricated when the lift off step was carried out in H$_2$O$_2$ (Fig. 1C and D).

The second approach (potential shock) was intended for fabricating membranes with a fully open bottom. In general, we grew nanotube layers for 1 h and then applied the "potential shock". After preliminary screening of the experimental conditions (results not shown), we found that key parameters for a lift off of the entire membrane (*i.e.*, no breakage) and complete bottom opening, are a 3 min-long potential shock of 160 V (Fig. 1E and F).

The resulting membrane layers were used to fabricate the photo-anodes as illustrated in (Fig. 2A). For the membrane transfer, FTO glasses were coated with a 2 μm-thick TiO$_2$ NP film and then the membranes were transferred onto this layer in both tube top up and tube top down configurations. The TiO$_2$ NP film granted good adhesion, mechanical stability and established good electric contact between tubes and FTO.

In order to investigate the effect of membrane configuration and tube morphology on the photovoltaic performance, the fabrication of the DSSCs was completed by sandwiching the dye-sensitized photo-anodes with Pt coated FTO glass as a counter electrode and by introducing the electrolyte, as described in the experimental section.

Fig. 2B shows the solid-state J-V curves of the solar cells while Fig. 2C summarizes their photovoltaic characteristics measured under AM 1.5 simulated solar light illumination. From these data, it is clear that the cell geometry, that is, a tube top up or tube top down configuration, is a key factor to enhance the solar cell efficiency. For example, in the case of tube membranes fabricated by potential shock, a significant cell efficiency enhancement from 7.60% up to 8.55% could be obtained by adopting a tube top down configuration. These findings can be ascribed to the following factors: *i)* tubes in such a configuration show better light absorption properties since light scattering given by the tube bottom is minimized (clearly, this refers to the case when a front side irradiation geometry is adopted) [17]; and *ii)* more efficient charge transport and collection are obtained when the tub top is in contact with the TiO$_2$ NP layer.

In particular, in order to confirm the latter assumption, we investigated by IMPS measurements the electron transport properties of DSSCs fabricated from membranes in the two different configurations (see data in Fig. 2D). A significant improvement of the electron transport was measured for the DSSCs that were fabricated with membranes in tube top down configuration. Precisely, such a configuration showed an electron transport time that is slower by one order of magnitude compared to that measured for the tube top up configuration.

Besides, we found that also the tube morphology largely affected the photovoltaic performance of the solar cells. In fact, solar cells fabricated from membranes with fully close bottom (detached by re-anodization and HF treatment) gave relatively low efficiency (7.08%). Instead, fully bottom-open membranes (detached by potential shock) led to higher solar cell efficiency (8.55%). Interestingly, a partial opening of the membrane tube bottom, which was obtained by re-anodization and $H_2O_2$ treatment, lead to a cell efficiency value (8.07%) that was intermediate compared to those measured for membranes with complete close and fully open bottom. In other words, the solar cell efficiency was shown to closely correlate to the extent of bottom opening of the tube membrane.

To evaluate if changes in crystallinity could be responsible for different cell efficiencies and transport properties, XRD patterns of the photo-anodes were collected after annealing at 500 °C. As shown in Fig. 3A, the $TiO_2$ NT membranes were observed to be composed of 100 % anatase phase, this regardless of fabrication method (*i.e.*, re-anodization or potential shock) and configuration of the membrane on the FTO slide (that is, tube top up or tube top down), indicating that the tube crystallographic properties cannot be responsible for the large difference in cell efficiency.

In order to better evaluate the key factors that affect the performance of the different membranes, data from Fig. 2A and B were re-plotted as shown in Fig. 3B, that is, compiling dye-loading, $J_{sc}$ and η in a single plot. This graph clearly shows that the amount of dye loaded on the tube membranes correlates very well with the $J_{sc}$ and η of the cells and, most importantly, all these parameters correlate in turn with the extent of bottom opening of the tube membrane. One thus may conclude that the open/close bottom structure of the membranes largely affect the ability of the sensitizer solution to penetrate the entire $TiO_2$ structure. In other words, a bottom open morphology (open porosity) allows for improved dye-loading, and this factor combined with effective electron transport of the tube top down configuration leads to a significant enhancement of the cell efficiency.

## 4. Conclusions

$TiO_2$ nanotube membranes detached from the substrates were used as photo-anodes to fabricate DSSCs. Different membrane lift off strategies were explored that allowed for fine control over the extent of bottom opening of the tubes. Besides, the membrane configuration in the DSSCs was shown to largely affect the charge transport and collection efficiency. Solar cells fabricated from fully bottom open membranes in tube top down configuration showed improved efficiencies compared to other morphologies and cell configurations. These results are ascribed to enhanced electron transport properties of tube top down configuration, and even more to the fact that the open porosity of tubes detached by potential shock leads to a specifically higher dye-loading.

**Acknowledgments**

The authors would like to acknowledge ERC, DFG and the Erlangen DFG cluster of excellence for financial support. Also the financial support from the Iran Ministry of Science, Research and Technology is gratefully acknowledged.

**Figure captions**

**Figure 1** - SEM pictures of: (A) typical TiO$_2$ nanotube array used for fabricating the membrane (the inset shows a top-view of the tube layer); (B) bottom close TiO$_2$ nanotube membrane prepared by re-anodization approach followed by detachment in aqueous HF solution (the inset shows a high-magnification of the tube bottom); (C and D) partially bottom open TiO$_2$ nanotube membrane prepared by re-anodization approach followed by detachment in aqueous H$_2$O$_2$ solution; (E and F) fully bottom open TiO$_2$ nanotube membrane prepared by a 3 min-long potential shock of 160 V.

**Figure 2** - (A) Sketch of a photo-anode consisting of 12 μm-thick TiO$_2$ NT membrane transferred in tube top down configuration onto FTO slide coated with 2 μm-thick TiO$_2$ NP film; (B) J-V curves measured under simulated solar light front-side illumination of DSSCs fabricated from 12 μm-thick TiO$_2$ NT membranes that were prepared by re-anodization and potential shock approach and transferred onto FTO in tube top up and tube top down configuration; (C) photovoltaic parameters of different solar cells; (D-E) IMPS measurements of DSSCs fabricated from tube membranes detached by potential shock and transferred onto FTO slides in different configurations.

**Figure 3** - (A) XRD patterns of different photo-anodes annealed in air at 500 °C for 1 h (heating/cooling rate of 30 °C min$^{-1}$); (B) dye-loading, short circuit current and power conversion efficiency of DSSCs fabricated from tube membranes that were transferred onto FTO glass in tube top down configuration. The SEM images show a close view of the tube bottom (the scale bar is of 200 nm).

**Figure 1**

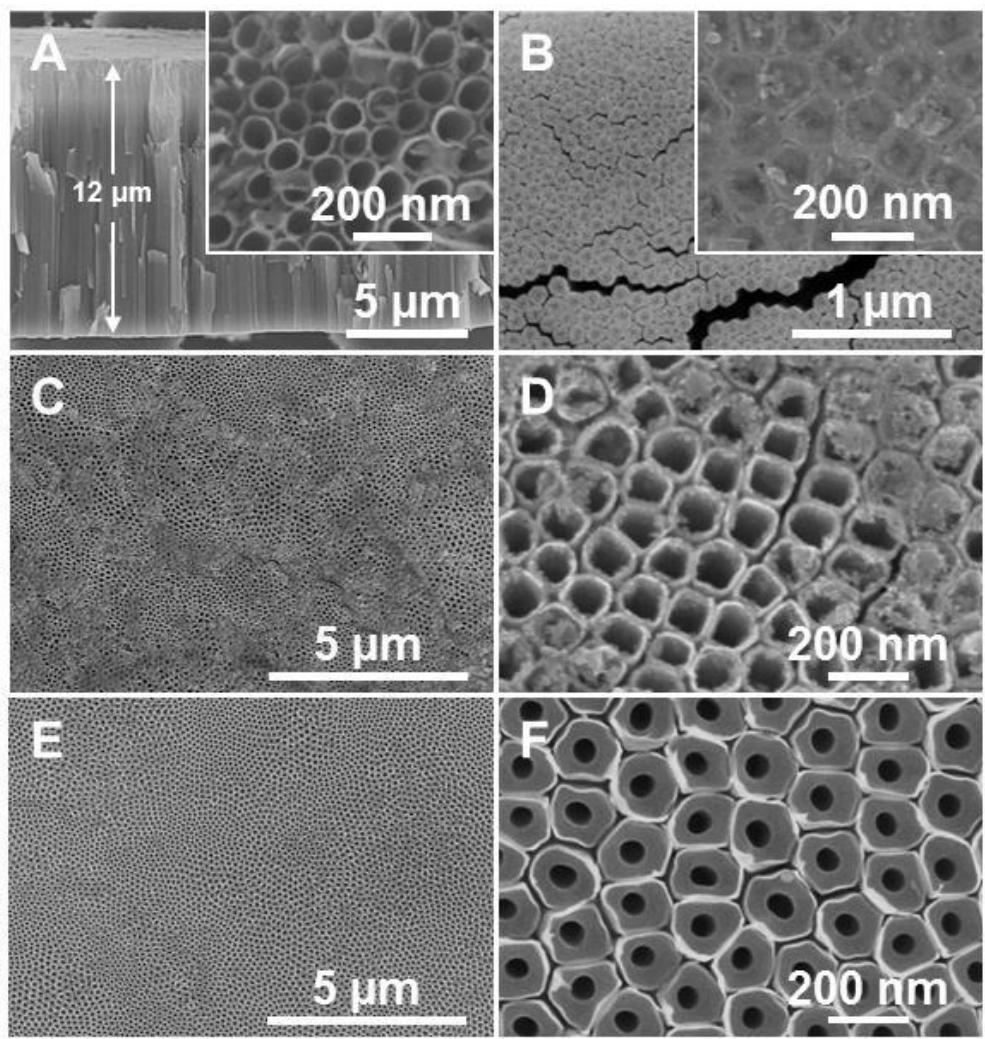

**Figure 2**

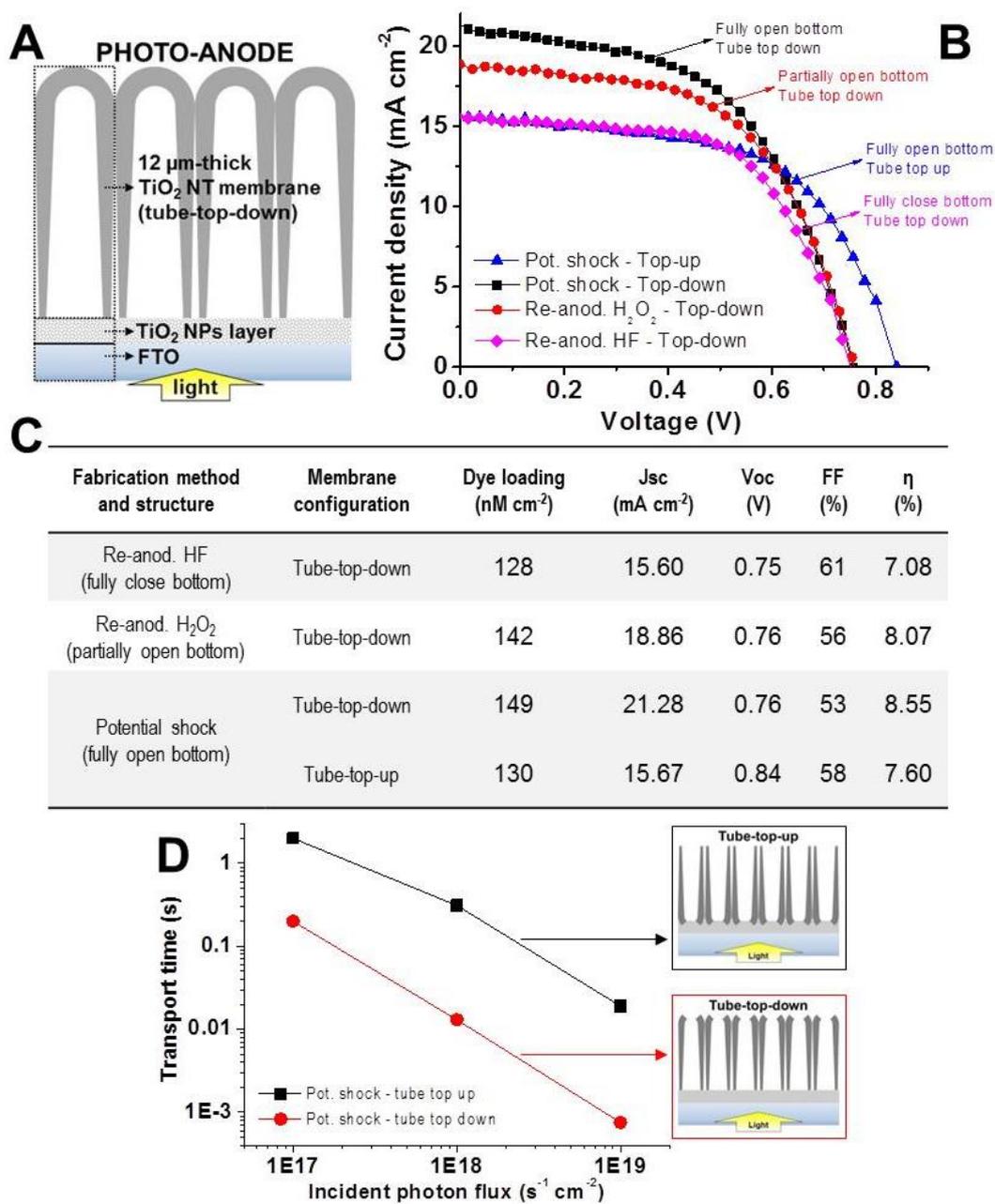

**Figure 3**

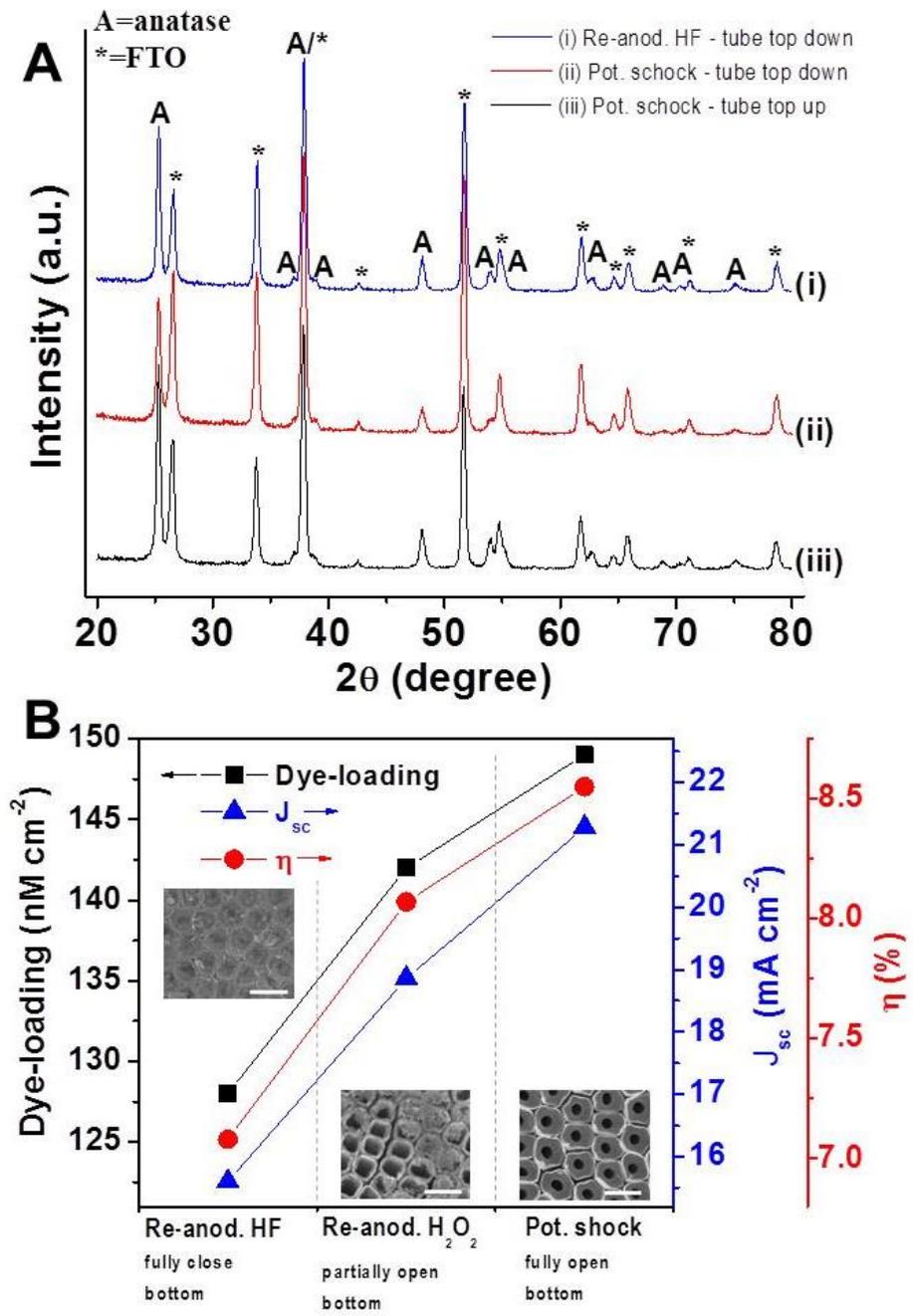